\documentclass[10pt,a4paper,twoside]{article}
\usepackage{epsfig}
\usepackage{baltlat6}
\usepackage{array}
\usepackage{dcolumn}
\usepackage{here}
\begin{document}
\ \
\vspace{0.5mm}
\setcounter{page}{19}
\vspace{8mm}

\titlehead{Baltic Astronomy, vol.\,18, 19--31, 2009}

\titleb{INTRINSIC COLOR INDICES AND LUMINOSITY SEQUENCES OF STARS IN THE
2MASS TWO-COLOR DIAGRAM}

\begin{authorl}
\authorb{V. Strai\v{z}ys}{1} and
\authorb{Romualda Lazauskait\.e}{1,2}
\end{authorl}

\begin{addressl}
\addressb{1}{Institute of Theoretical Physics and Astronomy, Vilnius
University,\\  Go\v{s}tauto 12, Vilnius LT-01108, Lithuania;
straizys@itpa.lt}
\addressb{2}{Department of Theoretical Physics, Pedagogical University,
Student\c{u} 39,\\ Vilnius, LT-08106, Lithuania; laroma@itpa.lt}
\end{addressl}

\submitb{Received: 2009 March 20; accepted: 2009 April 20}

\begin{summary} Intrinsic sequences of luminosity V and III stars in the
$J$--$H$ vs.  $H$--$K_s$ diagram of the 2MASS system are determined
using about 1250 unreddened and dereddened stars.  Intrinsic color
indices for MK spectral classes are tabulated and compared with the
results in other {\it JHK} systems.
\end{summary}


\resthead{Intrinsic color indices and sequences in the 2MASS
two-color diagram} {V.  Strai\v{z}ys, R. Lazauskait\.e}

\sectionb{1}{INTRODUCTION}

Due to its ability to identify many types of stars, all-sky coverage and
a faint limiting magnitude, the 2MASS {\it JHK}$_s$ system has become
one of the most popular photometric systems.  However, until now the
accuracy of intrinsic colors in the system is still insufficient.  The
users of the 2MASS system usually apply intrinsic colors transformed
from the Glass {\it JHK}$_s$ system (Bessell \& Brett 1988) with the
equations given by Carpenter (2001).  However, these equations are
linear, and they may not be valid for all types of stars, especially, at
the edges of the temperature scale, or for stars of different
luminosities, metallicities and peculiarities.  Consequently, a direct
determination of the intrinsic colors of stars in the 2MASS system is
desirable.

A direct calibration of color indices of 2MASS stars in spectral and
luminosity classes is not trivial.  Most stars with the best known
values of physical parameters and spectral types were too bright for
2MASS photometry.  According to Skrutskie et al.  (2006), images of
stars brighter than $K_s$\,=\,4.5 mag were saturated, and their
uncertainties are of the order of 0.2--0.4 mag.  On the other hand, most
of fainter stars, especially of early types or giants and supergiants,
are affected by interstellar reddening.  For their dereddening a
knowledge of the slopes of interstellar reddening lines and color
excesses is essential.  These data, however, are not always known with
sufficient accuracy.

Recently, Strai\v{z}ys, Corbally \& Laugalys (2008) have determined mean
intrinsic color indices ($J$--$H$)$_0$ and ($H$--$K_s$)$_0$ in the 2MASS
system of O8, B5.5 and B8.5 spectral classes by dereddening stars
with small color excesses.  In the present work we extend the
investigation to main-sequence stars of all spectral classes and to
late-type giants.  To obtain the mean intrinsic lines of luminosity V
and III stars in the $J$--$H$ vs.\,$H$--$K_s$ diagram we use both
unreddened stars and stars with small color excesses.

The $J$, $H$ and $K_s$ magnitudes are taken from the 2MASS database.  We
used only those stars which have the accuracy of magnitudes $\leq$\,0.05
mag.  All stars with magnitudes $J$, $H$ or $K_s$ brighter than 4.5 mag
(i.e., with saturated images) were rejected.

\sectionb{2}{MAIN-SEQUENCE STARS}

For the determination of intrinsic color indices of main-sequence stars
we tried to use, where possible, unreddened stars with reliable spectral
classification available.  One of such lists was published by Gray et
al.  (2003), covering many F, G and K dwarfs within 40 pc from the Sun.
In the same spectral range we used stars from the three open clusters --
Praesepe, Pleiades and M\,39 -- selecting the main-sequence stars
unsaturated in 2MASS photometry.  In the clusters, Am and Ap stars as
well as visual and spectroscopic binaries were rejected.  The Praesepe
stars were considered to be unreddened, but for the Pleiades and M\,39
stars the $J$--$H$ and $H$--$K_s$ colors were slightly corrected for
reddening.  Their spectral types and $B$--$V$ color indices were taken
from the WEBDA database.
\footnote{~http://univie.ac.at/webda/navigation.html}

The Pleiades also contain a good sample of B and A stars, but most of
these B stars were too bright for precise 2MASS photometry.  Thus, the
B-star sample was supplemented by members of the Orion OB1 association
of spectral classes B0--B2 from the Humphreys \& McElroy (1984) list and
of spectral classes B5--A2 from \v{C}ernis et al.  (1998).
Additionally, B-type stars of the association/cluster Collinder 121 with
the data from the WEBDA database were used.  The Orion association and
Collinder 121 stars were dereddened individually, their $E_{B-V}$ mostly
are $\leq$\,0.10.  To increase the number of stars, we added some
B-stars of luminosity IV since there is no evidence that they differ
from luminosity V in the near infrared.

O-type stars of luminosities V--III having little interstellar reddening
were selected from the catalog published by Ma\'iz-Apell\'aniz et al.
(2004).  From these, 16 stars of spectral classes O6.5--O9.5 have color
excesses $E_{B-V}$\,$\leq$\,0.25 and 9 stars of classes O3--O6 have
$E_{B-V}$ in the range 0.28--0.39.  Magnitudes $J$, $H$ and $K_s$
for these stars were taken from the original 2MASS catalog since in the
Ma\'iz-Apell\'aniz et al. catalog they have systematic differences.
For dereddening their $J$--$H$ and $H$--$K_s$ color indices we used
their $B$--$V$ from the above mentioned catalog and the intrinsic color
indices $(B-V)_0$ = --\,0.315.

M-type dwarfs, mostly of early subclasses, were taken from the Praesepe
and Pleiades clusters (Adams et al. 2002, Table 4).  More M-dwarfs of
the early subclass group (M0--M5.5 V) were taken from Kirkpatrick et al.
(1991) and Henry et al.  (1994).  Late M-dwarfs (M6--M9.5) were taken
from Kirkpatrick et al.  (1991), Gizis et al.  (2000) and Liebert \&
Gizis (2006).  A few K-dwarfs from Kirkpatrick et al.  (1991) were added
to the Gray et al.  (2003) list.  A sample of L-type dwarfs, which
includes 83 stars, was selected from Gizis et al.  (2000), Liebert \&
Gizis (2006), Sheppard \& Cushing (2009) and Kirkpatrick et al.  (2008).


\vbox{
\centerline{\psfig{figure=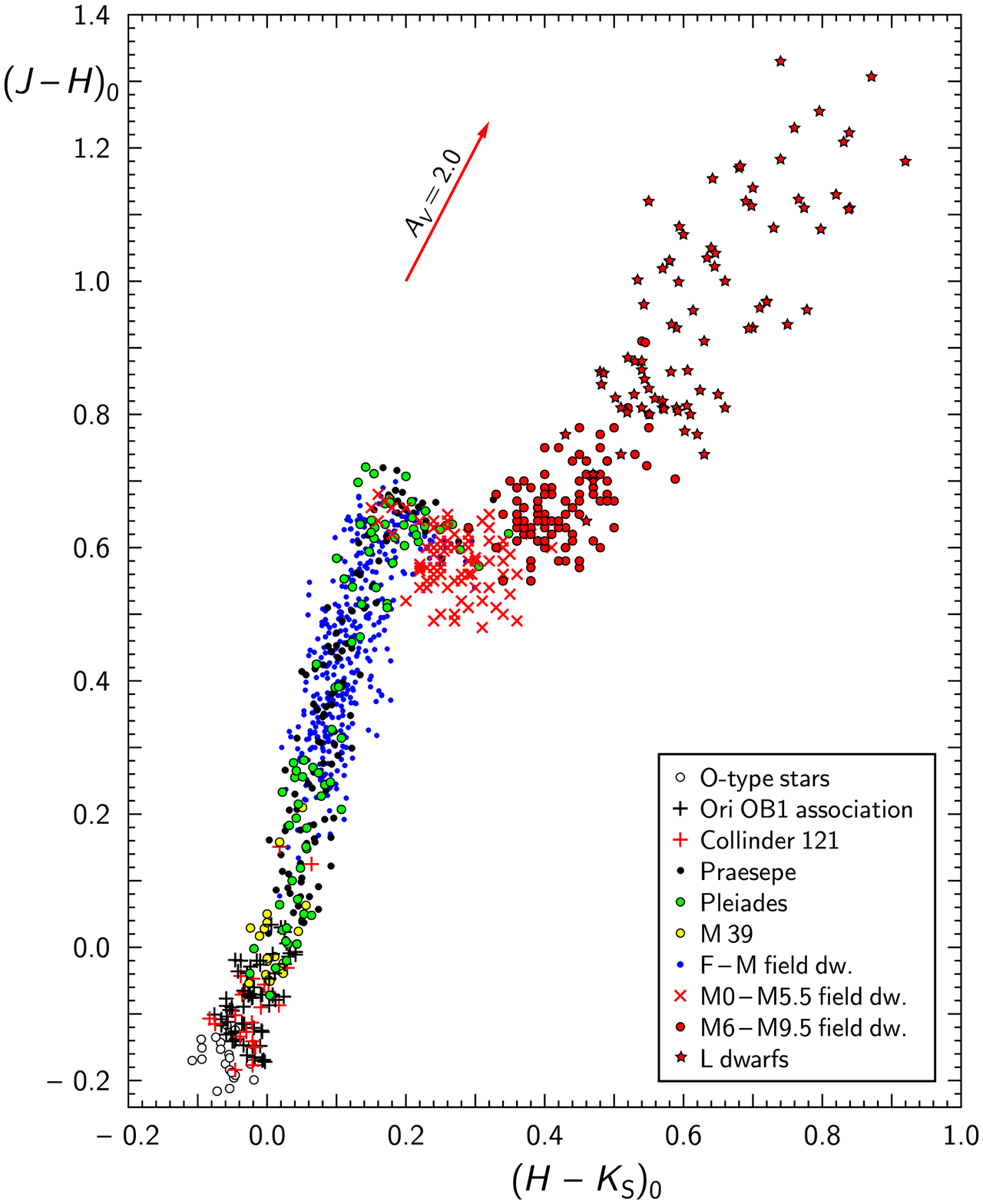,width=125mm,angle=0,clip=}}
\vspace{3mm}
\captionb{1}{The $J$--$H$ vs.\,$H$--$K_s$ diagram for unreddened or
dereddened main-sequence stars. The symbols are explained in the insert.}
}


\begin{figure}[!t]
\centerline{\psfig{figure=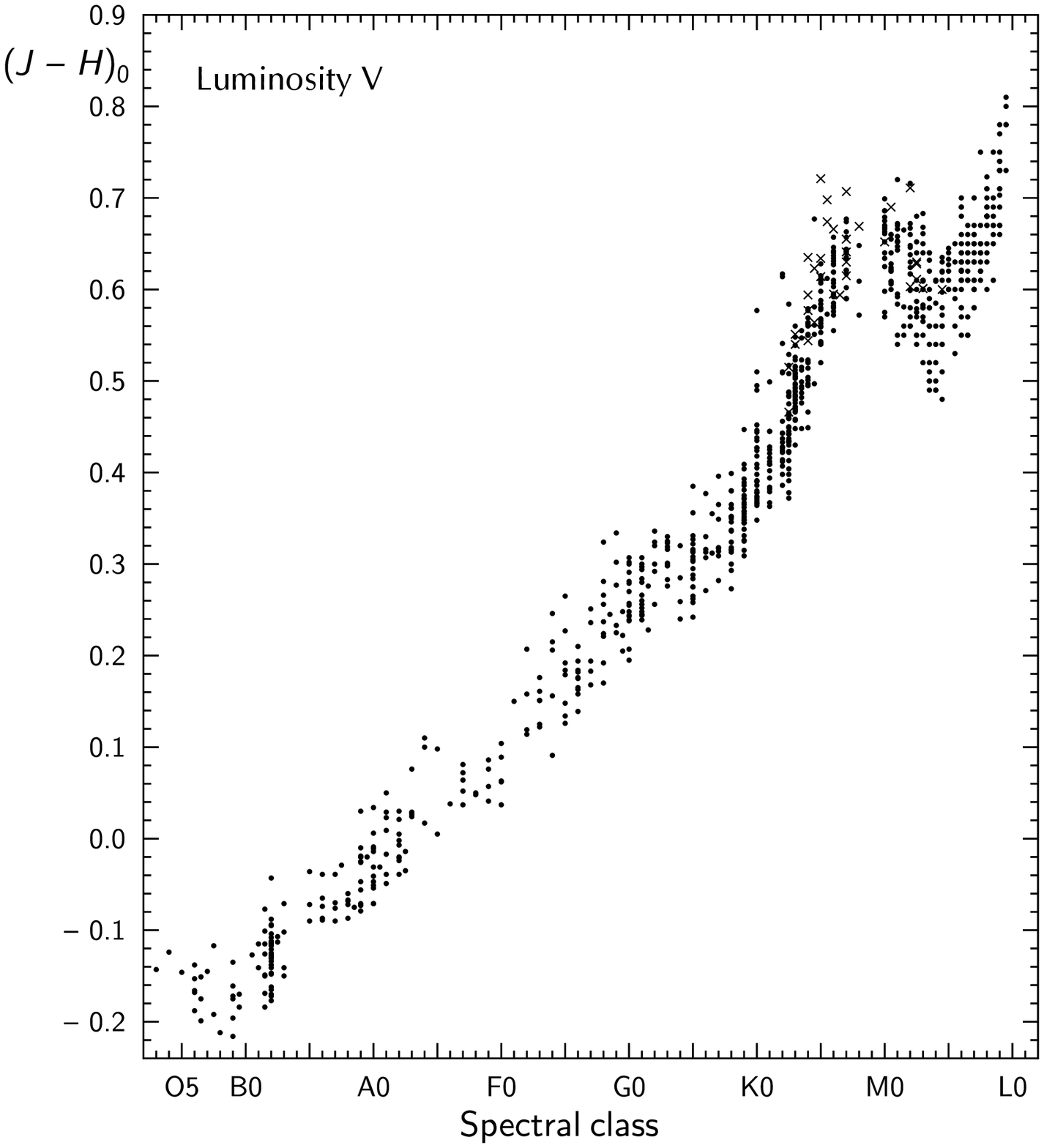,width=120mm,angle=0,clip=}}
\vspace{.2mm}
\captionb{2}{Intrinsic color indices $(J-H)_0$ of the main-sequence
stars plotted as a function of spectral class. Crosses denote stars
with chromospheric emission lines.}
\end{figure}
\vspace{3mm}

\begin{figure}[!t]
\centerline{\psfig{figure=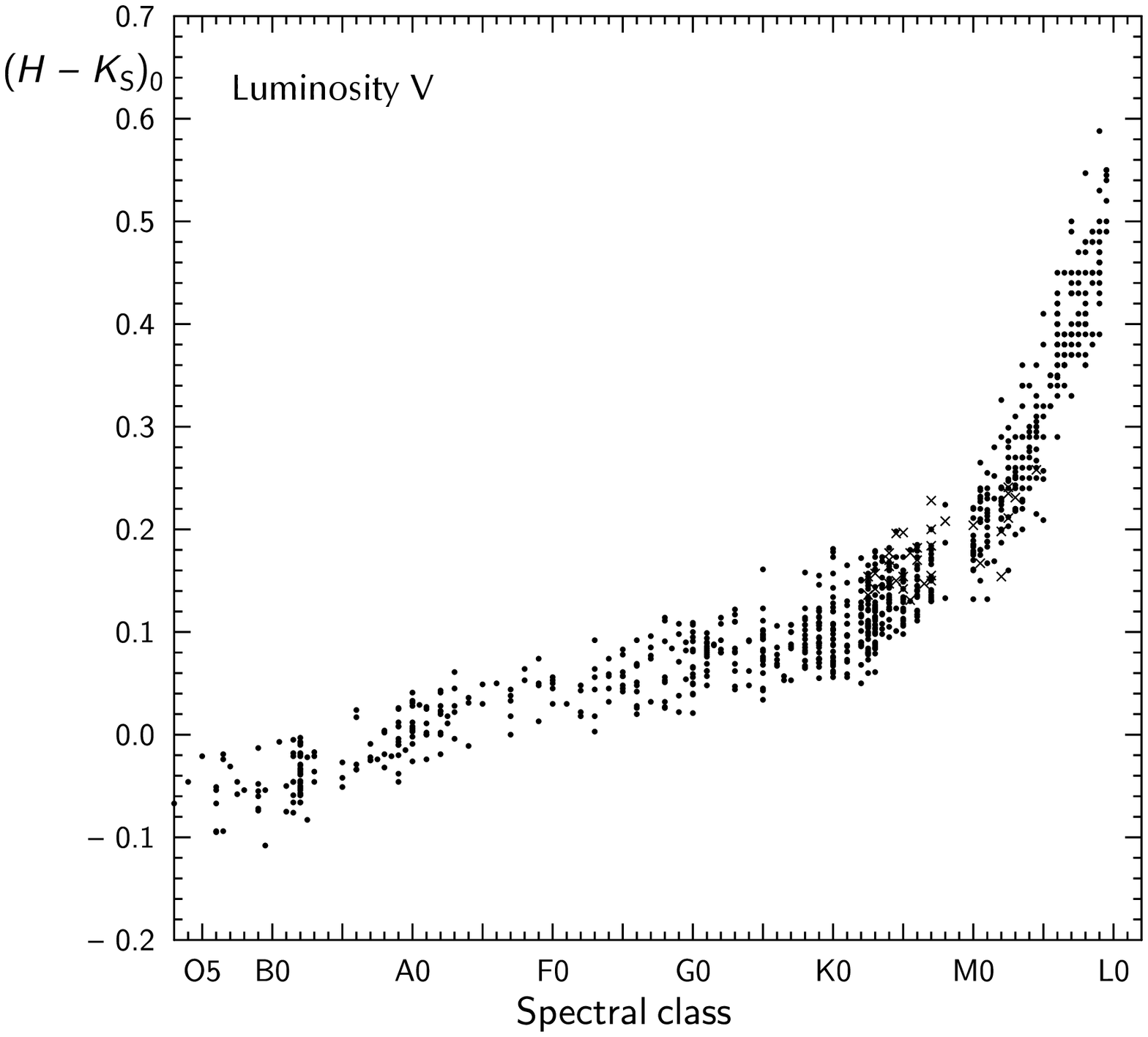,width=120mm,angle=0,clip=}}
\vspace{.2mm}
\captionb{3}{Intrinsic color indices $(H-K_s)_0$ of the main-sequence
stars plotted as a function of spectral class. Crosses denote stars
with chromospheric emission lines.}
\end{figure}

Intrinsic color indices ($J$--$H$)$_0$ and ($H$--$K_s$)$_0$ for the
selected 980 main-sequence stars are plotted in Figure 1. Different
groups of stars are shown by different symbols explained in the insert.
A few stars deviate considerably from the general
dependence; such outliers were not plotted.  The remaining stars form a
rather broad band with the $H$--$K_s$ amplitude increasing from about
0.1 mag for O to K stars up to 0.2 mag for M dwarfs of late subclasses.
A similar diagram for M and L dwarfs (for a somewhat different sample of
stars) was plotted by Kirkpatrick et al.  (2000).  The dispersion of
stars in the diagram mostly is the result of the observational errors of
color indices, since $\sigma$ values of color indices for some stars can
be $\pm$\,0.07 mag.  Additional effects contributing to this dispersion
can be differences in age and chromospheric activity, the presence of
warm circumstellar gas and dust disks or envelopes, unknown stellar or
substellar companions, differences in chemical composition, etc.  We
will discuss these possible effects in more detail.

It is known that up to 20\,\% of field K--M2 dwarfs have active
chromospheres and exhibit emission components of H$\alpha$, Ca\,H and K
and other lines.  In young clusters with ages $<$\,100 Myr (like the
Pleiades) almost all stars cooler than K4 show chromospheric activity.
In the Hyades and Praesepe this happens at spectral class M2. The
majority of stars of subclass M4 and cooler, both in the field and in
clusters, are flare stars with strong and variable emission lines.  In
Figure 1, late K and early M dwarfs lie at the upper edge of the main
sequence belt.  The most outstanding examples are the chromospherically
active Pleiades and Praesepe stars located at the top of the belt with
$J$--$H$\,$\approx$\,0.7.  Active M dwarfs of cooler subclasses with
strong H$\alpha$ emission do not show increase in their $J$--$H$.

Flat circumstellar dust disks are present around main-sequence stars of
spectral classes A, F, G and K (Habing et al. 2001).  Although for
M-dwarfs the results are discordant (Plavchan et al. 2005; Riaz et al.
2006), at least part of them also hold warm dust and debris disks.
Such disks create excesses of infrared flux, contributing to the $J$,
$H$ and $K_s$ magnitudes.

To estimate the metallicity effect, on the $J$--$H$ vs.\,$H$--$K_s$
diagram we plotted a few F-G-K subdwarfs with [Fe/H]\,$\approx$\,--2.
All of them lie in one sequence, together with dwarfs of solar
metallicity.  However, it is known that metal-deficient M-dwarfs (or
subdwarfs) in this diagram deviate down from their metal-rich
counterparts (see Leggett 1992 and references therein).

In Figures 2 and 3 we plot color indices $J$--$H$ and $H$--$K_s$ against
spectral classes without differenciating stars into physical groups.
Instead, the known emission-line stars of K- and early M-subclasses with
active chromospheres are plotted as crosses.  It is evident that these
stars exhibit larger color indices $J$--$H$ than non-emission stars.  A
similar tendency (however milder) is noticeable for $H$--$K_s$.  The
mean intrinsic color indices for various spectral classes were
calculated by the least-square method.  After plotting them as a
function of spectral class a smooth curve was drawn through the points
and the resulting color indices were read out from the curve.  The
results are given in Table 1. The standard deviations for most of these
indices are 0.02--0.05 mag.

\begin{table}[!t]
\vbox{\small\tabcolsep=3pt
\begin{center}
\centerline{\baselineskip=9pt {\smallbf Table 1.}{\small\ 2MASS intrinsic color indices
for luminosity V stars. \lstrut}}
\begin{tabular}{cD..{4.4}D..{4.4}|cD..{4.4}D..{4.4}}
\hline
\multicolumn{1}{c}{Sp. type\hstrut}&
\multicolumn{1}{c}{$(J-H)_0$}&
\multicolumn{1}{c|}{$(H-K_s)_0$}&
\multicolumn{1}{c}{Sp. type}&
\multicolumn{1}{c}{$(J-H)_0$}&
\multicolumn{1}{c}{$(H-K_s)_0$\lstrut} \\
\hline
\hstrut
O5\,V   &    -0.17  &  -0.06   &       K1\,V     &      0.415  &    0.105 \\
B0\,V     &    -0.15  &  -0.05   &     K2\,V    &       0.44  &   0.11   \\
B2\,V     &    -0.13  &  -0.04   &     K3\,V    &       0.49  &   0.12   \\
B5\,V     &    -0.09  &  -0.03   &     K4\,V    &       0.54  &   0.135  \\
B8\,V     &    -0.06  &  -0.02   &     K5\,V    &       0.59  &   0.145  \\
A0\,V     &    -0.04  &   0.01   &     K6\,V    &       0.62  &   0.15   \\
A2\,V     &    -0.01  &   0.02   &     K7\,V    &       0.64  &   0.16   \\
A5\,V     &     0.03  &   0.035  &     M0\,V    &       0.65  &   0.18   \\
A7\,V     &     0.06  &   0.04   &     M1\,V    &       0.64  &   0.20   \\
F0\,V     &     0.09  &   0.045  &     M2\,V    &       0.63  &   0.23   \\
F2\,V     &     0.12  &   0.05   &     M3\,V    &       0.60  &   0.245  \\
F5\,V     &     0.18  &   0.055  &     M4\,V    &       0.56  &   0.28   \\
F8\,V     &     0.23  &   0.06   &     M5\,V    &       0.60  &   0.34   \\
G0\,V     &     0.26  &   0.07   &     M6\,V    &       0.63  &   0.38   \\
G2\,V     &     0.28  &   0.075  &     M7\,V    &       0.65  &   0.41   \\
G5\,V     &     0.30  &   0.08   &     M8\,V    &       0.67  &   0.44   \\
G8\,V     &     0.34  &   0.09   &     M9\,V    &       0.72  &   0.47   \\
K0\,V     &     0.39  &   0.10   &              &             &  \lstrut  \\
\hline
\end{tabular}
\end{center}
\begin{center}
\parbox{90mm}{\baselineskip=9pt {\smallbf Table 2.}{\small\ Open clusters
with red giants used for the determination of the intrinsic colors. \lstrut}}
\begin{tabular}{lD..{4.4}D..{4.4}D..{4.4}}
\hline
\multicolumn{1}{l}{\dnnn{Cluster}\hstrut}&
\multicolumn{1}{c}{\dnnn{$E_{B-V}$}}&
\multicolumn{1}{c}{\dnnn{Number of giants}}&
\multicolumn{1}{c}{Number of}\\ & & & \multicolumn{1}{c}{\uppp{MK
spectra}}\\
\hline
\noalign{\vskip1mm}
NGC\,2099 = M\,37    &    0.30      &          29     &        4 \\
NGC\,2506            &    0.08      &          34     &       -  \\
NGC\,2682 = M\,67    &    0.06      &          53     &       10  \\
NGC\,6067            &    0.38      &          12     &        5  \\
NGC\,6791*           &    0.12      &          73     &        3  \\
NGC\,6819            &    0.24      &          47     &        -  \\
NGC\,7789            &    0.22      &         123     &        6  \\
\hline
\noalign{\vskip1mm}
\multicolumn{4}{c}{\small
* Only three stars of this cluster were used, see the text.\hstrut}\\
\end{tabular}
\end{center}
}
\vspace{-3mm}
\end{table}

\sectionb{3}{LATE-TYPE GIANTS}

The determination of intrinsic 2MASS colors for red giants is more
complicated since most of the unreddened field giants are too bright and
the accuracy of their colors is rather low due to saturation.  The
visual magnitude, at which the saturation of {\it JHK} images takes
place, depends strongly on spectral type:  $V$\,$\approx$\,6.0 mag for
spectral type G5\,III, 7.5 mag for K5\,III and 10 mag for M5\,III.  If
the stars are reddened or, in case of late M subclasses, have warm dust
envelopes, the corresponding $V$ magnitudes are even fainter.


\begin{figure}[!t]
\centerline{\psfig{figure=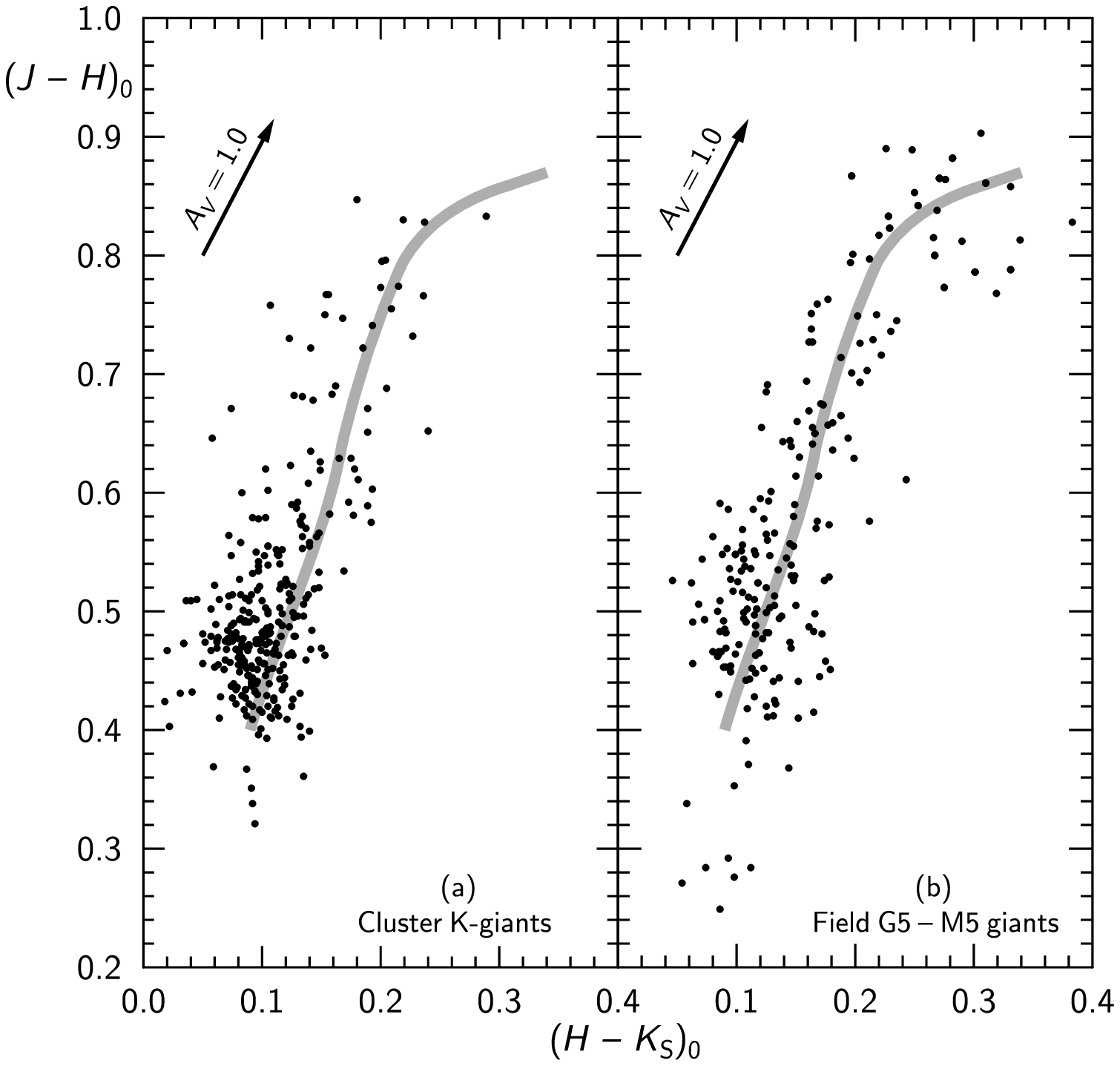,width=100mm,angle=0,clip=}}
\vspace{.2mm}
\captionb{4}{The $J$--$H$ vs.\,$H$--$K_s$ diagram for unreddened or
dereddened giants of spectral classes G5\,III and cooler.  Panel (a)
shows the open cluster members and panel (b) -- field giants from the
vicinity of NGP.}
\end{figure}

Field giants can be used for determining the intrinsic colors only if we
estimates of their color excesses.  Interstellar reddening is usually
rather low at high Galactic latitudes, especially at the Galactic poles,
where color excesses in the infrared may be neglected.  In all other
cases, color indices should be dereddened.  To do this, we must have
knowledge of MK spectral types and $B$--$V$ color indices.  Since
accurate photometry for many field stars is not available, their
reddenings cannot be estimated individually.  Therefore, we decided to
use red giants which are either located at high Galactic latitudes or
known to be members of old- or medium-age open clusters with color
excesses estimated from photometry in the {\it UBV} or other systems.
To reduce dereddening errors, we used only clusters with
$E_{B-V}$\,$\leq$\,0.4, these are listed in Table 2. However, in the
case of NGC\,6791, only three K-giants with available MK types have been
taken, since other giant-branch stars exhibit large scatter in the
cluster's color-magnitude diagram.

However, the majority of red giants in open clusters cover a relatively
narrow range of spectral classes, from G8 to K5.  More G--K--M giants
with small interstellar reddening can be found near the Galactic poles.
It is known that $E_{B-V}$ at the NGP is only about 0.03 mag up to the
Galaxy edge (see the review by Strai\v{z}ys 1992).  Thus, we are safe to
expect that for field giants in this direction $E_{J-H}$ will be less
than 0.01 mag, and $E_{H-K_s}$ less than 0.005 mag.  We selected about
170 late-type giants with MK types and unsaturated 2MASS images in the
vicinity of NGP from the Upgren (1960, 1962) catalogs.  Upgren's
spectral types were determined from low dispersion objective-prism
plates, so their accuracy is not high.  About 50 stars were added from
Helfer \& Sturch (1970), Schild (1973) and from the Skiff (2009)
compilation.  All of the stars used are at Galactic latitudes
$b$\,$>$\,65\degr.  Metallicity effect in the $J$--$H$ vs.\,$H$--$K_s$
diagram was verified by plotting dereddened field metal-deficient giants
with [Fe/H] between --2.0 and 3.0.  No systematic deviations from the
sequence of solar composition giants were found.

Almost all giants cooler than M5 are long-period variables of
semiregular or Mira types with variable magnitudes, color indices and
spectral classes.  Most of them are too bright in the near infrared, and
their magnitudes in the 2MASS catalog are of low accuracy.  Thus
their positions in the $J$--$H$ vs.\,$H$--$K_s$ diagram are uncertain
due to both variability and the saturated images. Therefore, in the
present investigation we limit ourselves only to giants of spectral
type M5\,III and earlier.

In Figure 4 (panels a and b) we show the 2MASS two-color diagram for 298
dereddened giants from open clusters and 280 field giants from the
vicinity of NGP, their spectral interval being from G5 to M5. For the
cluster giants we see the well-known clustering around $J$--$H$\,=\,0.46
and $H$--$K_s$\,=\,0.09, corresponding to the red clump giants (RCGs) in
their color-magnitude diagrams.  We shall return to RCGs below in this
section.  Since the NGP stars do not represent a specific volume of
space, we should not pay attention to the distribution of points in
Figure 4b.  These stars were chosen just to cover the whole
giant-sequence length between spectral classes G5 and M5.

MK spectral types are available for 28 stars in our open cluster sample
(most of luminosity class III, a few of II/III and II classes) and for
280 field stars.  Figures 5 and 6 show plots of color indices of these
stars as a function of spectral class.  The mean intrinsic colors for
different MK types given in Table 3 were determined by combining smooth
lines drawn through the middle of scattered points in Figures 4, 5 and
6. Due to large scatter the mean intrinsic colors are rather uncertain,
with their standard deviations being as large as 0.04--0.06 mag.  This
scatter is a result of errors in spectral classes and 2MASS color
indices, as well as due to unresolved binarity and peculiarity of the
stars.  The most uncertain are the $J$--$H$ colors for the giants of
spectral type G5\,III.  All of the stars lying in Figure 4b lower than
$J$--$H$ = 0.35 were classified by Upgren (1962) as G5--G6 giants.  We
do not exclude that some of them can be subgiants or even dwarfs.  In
this case the mean $J$--$H$ color of true G5 giants can be larger than
the value given in Table 3.

Some differences between the two panels of Figure 4 can be noticed.  For
some reason, the cluster K-giants, including the RCGs, show a small
systematic deviation to the left from the grey curve and a larger
scatter in the range of K3--K5 subclasses.  This cannot be explained by
dereddening errors since the reddening line shown in the diagrams of
Figure 4 is approximately parallel to the sequence of giants.

Special attention should be paid to the intrinsic colors of the red
clump giants, i.e., stars burning helium in the cores and hydrogen in
the shells.  In Strai\v{z}ys et al.  (2008) the values of ($J$--$H$)$_0$
= 0.502 and ($H$--$K_s$)$_0$ = 0.136 were obtained by averaging color
indices of seven RCGs of the cluster M\,67, which were considered


\vbox{
\centerline{
\epsfig{figure=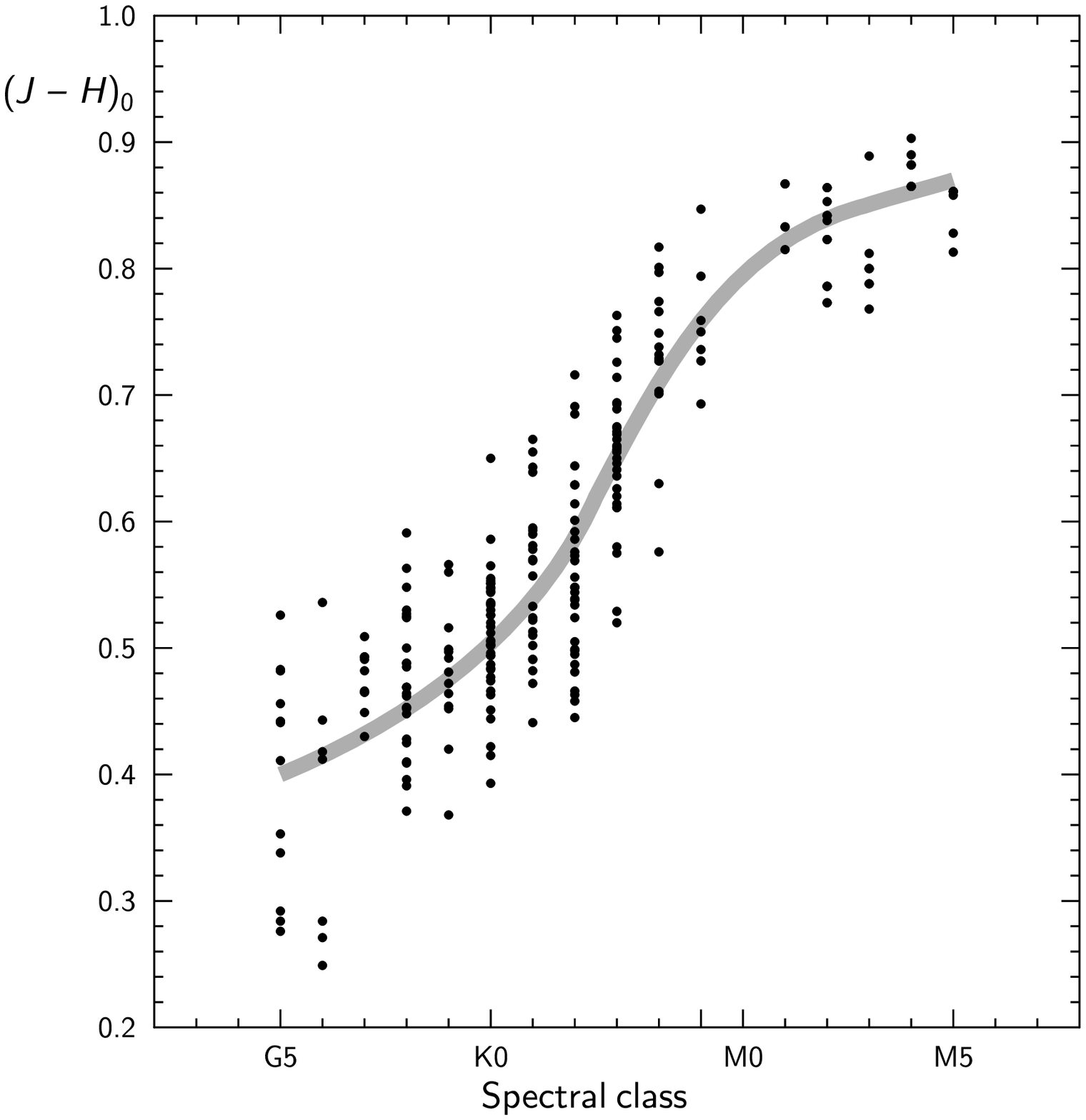,width=95mm,clip=}}
\vspace{-.5mm}
\captionb{5}{Intrinsic color indices $(J-H)_0$ of luminosity III stars
plotted as a function of spectral class. The gray curve corresponds to
the data of Table 3.}
\vskip6mm

\centerline{
\epsfig{figure=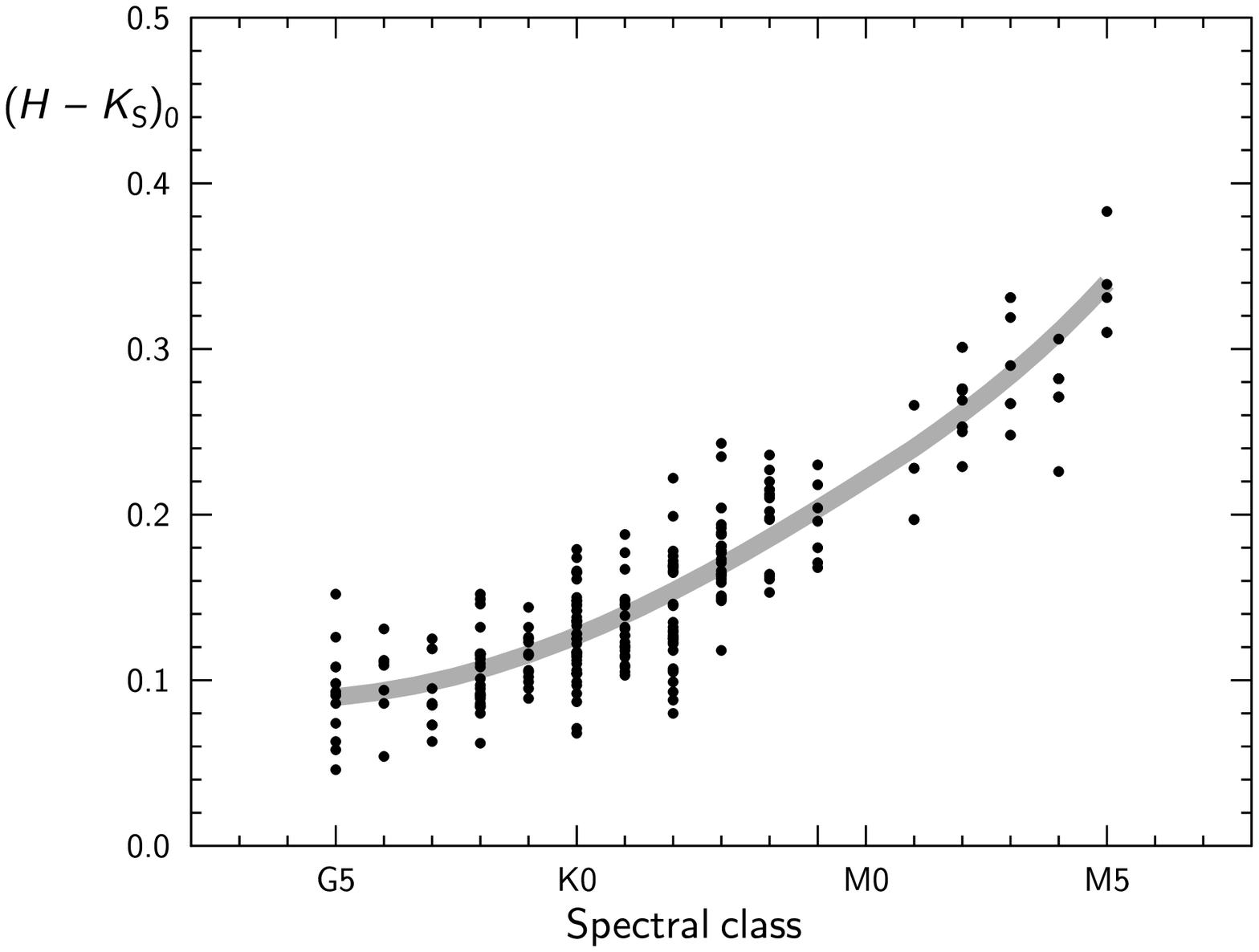,width=95mm,clip=}}
\vspace{-.5mm}
\captionb{6}{Intrinsic color indices $(H-K_s)_0$ of luminosity III stars
plotted as a function of spectral class. The gray curve corresponds to
the data of Table 3.}
\vskip3mm
}


\begin{table}[!h]
\vbox{\small\tabcolsep=3pt
\begin{center}
\centerline{\baselineskip=9pt {\smallbf Table 3.}{\small\ 2MASS intrinsic color indices
for luminosity III stars. \lstrut}}
\begin{tabular}{cD..{4.4}D..{4.4}|cD..{4.4}D..{4.4}}
\hline
\multicolumn{1}{c}{Sp. type\hstrut}&
\multicolumn{1}{c}{$(J-H)_0$}&
\multicolumn{1}{c|}{$(H-K_s)_0$}&
\multicolumn{1}{c}{Sp. type}&
\multicolumn{1}{c}{$(J-H)_0$}&
\multicolumn{1}{c}{$(H-K_s)_0$\lstrut} \\
\hline
\hstrut
G5\,III  &   0.40      & 0.09      &    K5\,III   &  0.76   &    0.205   \\
G8\,III  &   0.45      & 0.105     &    M0\,III   &  0.80   &    0.22    \\
K0\,III  &   0.50      & 0.125     &    M1\,III   &  0.82   &    0.24    \\
K1\,III  &   0.54      & 0.14      &    M2\,III   &  0.84   &    0.26    \\
K2\,III  &   0.59      & 0.155     &    M3\,III   &  0.85   &    0.285   \\
K3\,III  &   0.65      & 0.17      &    M4\,III   &  0.86   &    0.31    \\
K4\,III  &   0.71      & 0.185     &    M5\,III   &  0.87   &    0.34    \\
\hline
\end{tabular}
\end{center}
}
\end{table}

\begin{figure}[!th]
\centerline{\psfig{figure=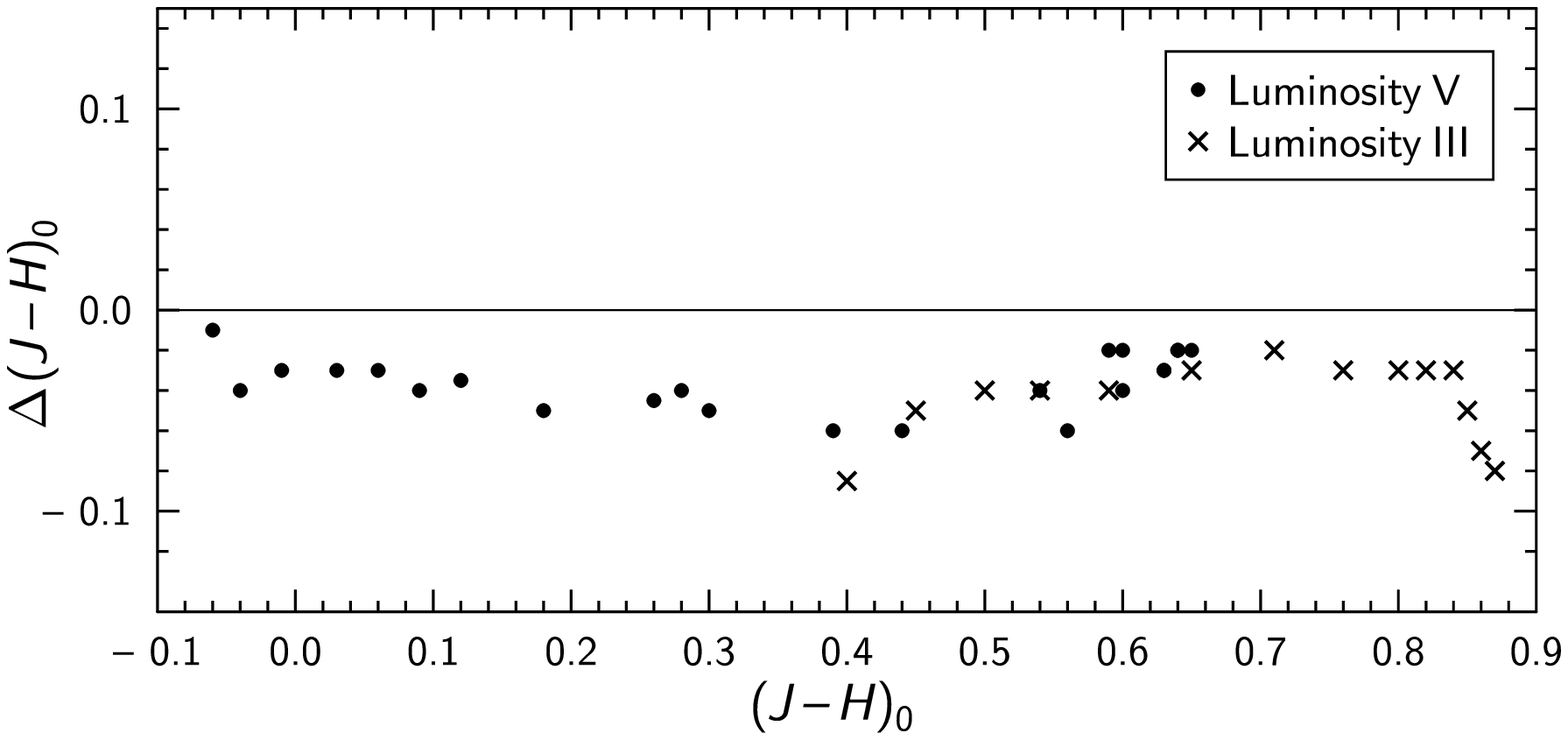,width=124mm,angle=0,clip=}}
\vspace{.2mm}
\captionb{7}{Differences between the intrinsic color indices $(J-H)_0$
given
in Tables 1 and 3 and those of Bessell \& Brett (1988) and Bessell
(1991).}
\vspace{8mm}
\centerline{\psfig{figure=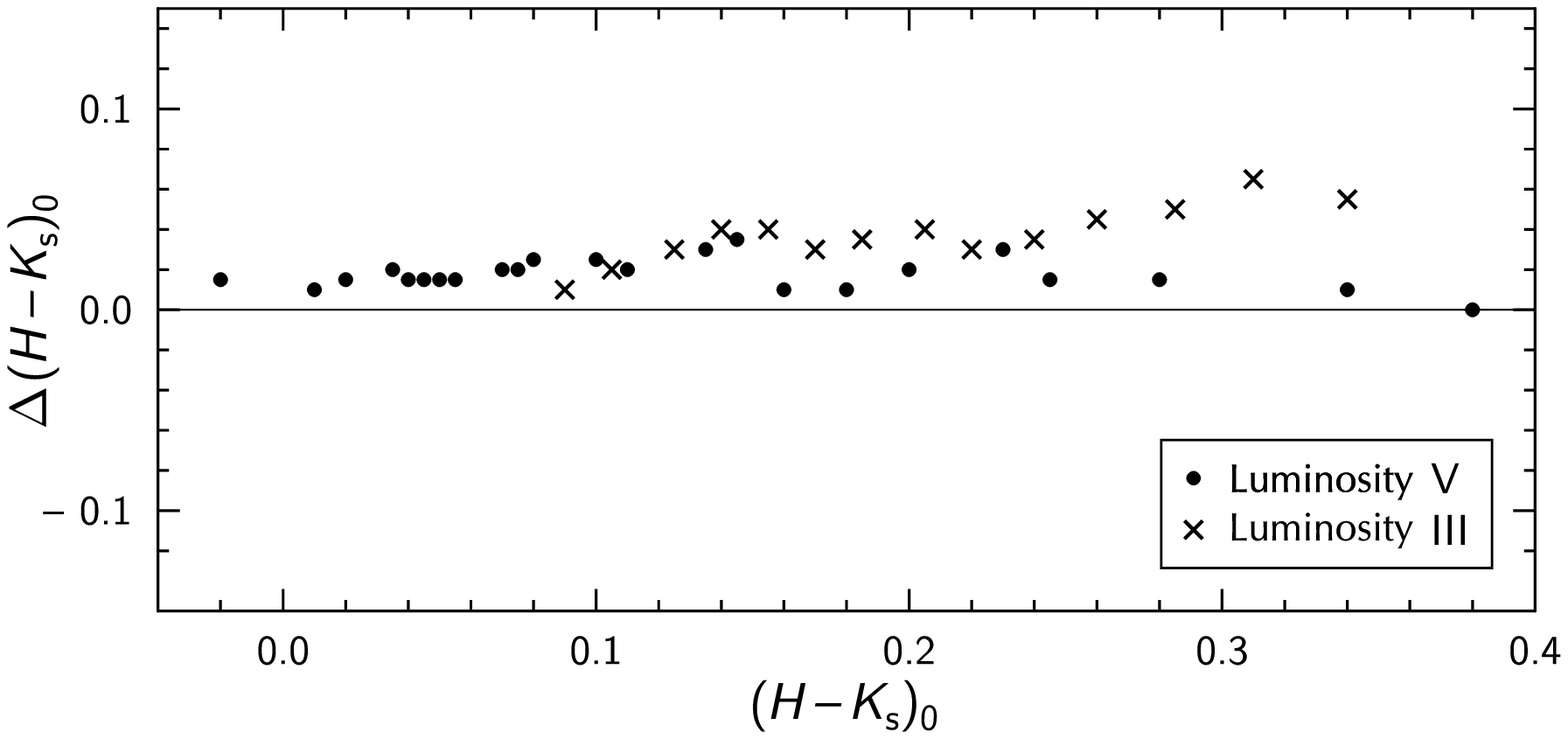,width=124mm,angle=0,clip=}}
\vspace{.2mm}
\captionb{8}{Differences between the intrinsic color indices $(H-K_s)_0$
given
in Tables 1 and 3 and those of Bessell \& Brett (1988) and Bessell
(1991).}
\end{figure}


\newpage

\noindent to be unreddened.  If little reddening of the
cluster is taken into account ($E_{B-V}$ = 0.06), the intrinsic values
of RCGs become 0.48 and 0.13.  Figure 4a shows that we have in the
present paper more than 100 stars in six clusters, which fall near the
RCG location.  Values of their $J$--$H$ colors concentrate within 0.44
and 0.48 and of $H$--$K_s$ within 0.06 and 0.12.  The mean values,
$J$--$H$ = 0.46 and $H$--$K_s$ = 0.09, are close to the intrinsic colors
of a mean G8\,III star.

\sectionb{4}{DISCUSSION AND CONCLUSIONS}

In this investigation we have determined the intrinsic sequences
of luminosity V and III stars in the $J$--$H$ vs.\,$H$--$K_s$ diagram of
the 2MASS photometric system using about 980 stars of the main sequence
and about 270 of late-type giants.  Based on stars with known spectral
classification, the intrinsic colors $(J-H)_0$ and $(H-K_s)_0$ for
different MK spectral types were determined.  To our knowledge, this is
the first direct calibration of the $J$--$H$ vs.\,$H$--$K_s$ diagram of
the 2MASS photometric system in MK types covering a broad range of
temperatures.  We find that the red clump giants of open clusters
concentrate at $J$--$H$ = 0.46 and $H$--$K_s$ = 0.09, but the spread of
points around this point is of the order of $\pm$\,0.03. The location of
 the clump center corresponds approximately to intrinsic
color indices for spectral class G8\,III (Table 3).

The range of spectral subclasses, estimated from color indices of RCGs
in different photometric systems, is from about G8 to K1.  According to
Grocholski \& Sarajedini (2002), the red clump giants in open clusters
have $J$--$K$ values of 0.6.  This is not very different from our value
for the clump, 0.55, but closer to spectral type K0\,III.  For red clump
giants in the solar neighborhood Alves (2000) finds $V$--$K$ in the
range 2.1--2.5, what corresponds to spectral types from G8\,III to
K1\,III (with the calibration from Strai\v{z}ys 1992).  The same range
of spectral subtypes is covered by 72\,\% of the RCGs from the Alves
list supplemented with spectral types from Simbad.  This spectral range
corresponds to $B$--$V$ between 0.9 and 1.1 in the well-known HR diagram
based on the {\it Hipparcos} data (Perryman et al. 1995, 1997).

In Figures 7 and 8 we compare the intrinsic color indices from Tables 1
and 3 with those for B8--K5 V and G5--M5 III stars from Bessell \& Brett
(1988), taking for K7--M7.5 V stars the corrected values from Bessell
(1991).  The intrinsic colors in these two papers have been determined
in a homogenized {\it JHK} system, named the Johnson-Glass system.  It
is evident that the differences between the two systems are non-linear
and multivalued.  By indirect comparison of the 2MASS and Bessell color
indices, Carpenter (2001) finds that the differences are almost constant
and color independent.  Zero point differences of the linear
transformation equations given by Carpenter, --0.045 for $J$--$H$ and
+0.028 for $H$--$K_s$, are in satisfactory agreement with Figures 7 and
8. However, in some ranges of colors, especially for M-type stars, there
are systematic deviations from the Carpenter equations, depending on
color and luminosity:  up to 0.04 mag in $\Delta$\,$J$--$H$ and 0.03 mag
in $\Delta$\,$H$--$K_s$.

The mean color indices for M dwarfs in the CIT {\it J,H,K} system were
also published by Leggett (1992), separately for the young disk, old
disk and halo objects.  Recently, Kirkpatrick et al.  (2000) and Liebert
\& Gizis (2006) investigated intrinsic colors for late M- and L-dwarfs
in the 2MASS system.  Our colors of M-dwarfs are in good agreement with
their results, despite the large scatter of stars used for averaging in
both studies.

One more possibility of determining the intrinsic colors is to
calculate  synthetic color indices for spectral energy distributions
of model atmospheres by convolving them with response functions of the
passbands.  However, our test calculations of color indices for the
Kurucz (2001) and Castelli \& Kurucz (2003) models show considerable
systematic differences between the synthetic and observational data,
especially for O--B and M stars.

For more exact determination of the intrinsic colors, observations in
the 2MASS filters of a set of bright stars with well-known physical
parameters are needed.  To avoid saturation, either much smaller
telescope or grey filters should be applied.

\thanks{We are thankful to Mathias Schultheis, Sta\-ni\-slava
Barta\v{s}i\={u}t\.e and Kazimieras Zdanavi\v{c}ius for
useful advices and Edmundas Mei\v{s}tas for the help in
preparing the paper.  The use of the 2MASS, SkyView, Gator,
Simbad and WEBDA databases is acknowledged.}

\References

\refb Adams J. D., Stauffer J. R., Skrutskie M. F. et al. 2002, AJ, 124,
1570

\refb Alves D. R. 2000, ApJ, 539, 732

\refb Bessell M. S. 1991, AJ, 101, 662

\refb Bessell M. S., Brett J. M. 1988, PASP, 100, 1134

\refb Carpenter J. M. 2001, AJ, 121, 2851

\refb Castelli F., Kurucz R. L. 2003, in {\it Modelling of Stellar
Atmospheres}, IAU Symp. 210, eds.  N. Piskunov, W. W. Weiss \& D. F.
Gray, ASP, p.\,A20\\ (http://kurucz.harvard.edu/grids/gridxxxodfnew)

\refb \v{C}ernis K., Barta\v{s}i\={u}t\.e S., Strai\v{z}ys V., Janulis
R. 1998, Baltic Astronomy, 7, 625

\refb Gizis J. E., Monet D. G., Reid I. N., Kirkpatrick J. D., Liebert
J., Williams R. J. 2000, AJ, 120, 1085

\refb Gray R. O., Corbally C. J., Garrison R. F., McFadden M. T.,
Robinson P. E. 2003, AJ, 126, 2048

\refb Grocholski A. J., Sarajedini A. 2002, AJ, 123, 1603

\refb Habing H. J., Dominik C., Jourdain de Muizon M. et al. 2001, A\&A,
365, 545

\refb Helfer H. L., Sturch C. 1970, AJ, 75, 971

\refb Henry T. J., Kirkpatrick J. D., Simons D. A. 1994, AJ, 108, 1437

\refb Humphreys R. M., McElroy D. B. 1984, {\it Catalogue of Stars in
Stellar Associations and Young Clusters}, Univ. of Minnesota, CDS
Strasbourg, Catalogue V/44

\refb Kirkpatrick J. D., Henry T. J., McCarthy D. W. 1991, ApJS, 77, 417

\refb Kirkpatrick J. D., Reid I. N., Liebert J. et al. 2000, AJ, 120,
447

\refb Kirkpatrick J. D., Cruz K. L., Barman T. S. et al. 2008, ApJ, 689,
1295

\refb Kurucz R. L. 2001,
http://kurucz.harvard.edu/grids/gridP00/fp00k2.pck

\refb Leggett S. K. 1992, ApJS, 82, 351

\refb Liebert J., Gizis J. E. 2006, PASP, 118, 659

\refb Ma\'iz-Apell\'aniz J., Walborn N. R., Galu\'e H. A., Wei L. H.
2004, ApJS, 151, 103

\refb Perryman M.\,A.\,C., Lindegren L., Kovalevsky J. et al. 1995,
A\&A, 304, 69

\refb Perryman M.\,A.\,C., Lindegren L., Kovalevsky J. et al. 1997,
A\&A, 323, L49

\refb Plavchan P., Jura M., Lipscy S. J. 2005, ApJ, 631, 1161

\refb Riaz B., Mullan D. J., Gizis J. E. 2006, ApJ, 650, 1133

\refb Schild R. E. 1973, AJ, 78, 37

\refb Sheppard S. S., Cushing M. C. 2009, AJ, 137, 304

\refb Skiff B. A. 2009, {\it General Catalogue of Stellar Spectral
Classifications}, Lowell\\ Observatory (available at CDS)

\refb Skrutskie M. F., Cutri R. M., Stiening R., Weinberg M. D. et al.
2006, AJ, 131, 1163

\refb Strai\v{z}ys V. 1992, in {\it Multicolor Stellar Photometry},
p. 146, Pachart Publishing House, Tucson, Arizona

\refb Strai\v{z}ys V., Corbally C. J., Laugalys V. 2008, Baltic
Astronomy, 17, 125

\refb Upgren A. R. 1960, AJ, 65, 644

\refb Upgren A. R. 1962, AJ, 67, 37

\end{document}